# Synchrotron radiation from a curved plasma channel laser wakefield accelerator

J.P. Palastro[1], D. Kaganovich[1], B. Hafizi[1], Y.-H. Chen[2], L.A. Johnson[1], J.R. Penano[1], M.H. Helle[1], and A.A. Mamonau[2]

[1] Naval Research Laboratory, Washington DC 20375-5346, USA
[2] Research Support Instruments, Lanham MD 20706 USA

**Abstract**

A laser pulse guided in a curved plasma channel can excite wakefields that steer electrons along an arched trajectory. As the electrons are accelerated along the curved channel, they emit synchrotron radiation. We present simple analytical models and simulations examining laser pulse guiding, wakefield generation, electron steering, and synchrotron emission in curved plasma channels. For experimentally realizable parameters, a ~2 GeV electron emits 0.1 photons per cm with an average photon energy of multiple keV.

I. Introduction

A high intensity ultrashort laser pulse propagating through plasma ponderomotively excites plasma waves [1-3]. The associated electric fields, far surpassing those of a conventional accelerator, can accelerate electrons to multiple GeV in distances as short as a centimeter [4-6]. The promise of harnessing these fields as a small-scale particle accelerator has led to a number of experimental demonstrations of and novel theoretical concepts for laser wakefield acceleration (LWFA) [4-12]. The strong transverse accelerations and large energy gains in LWFA have also enabled small-scale X-ray source development, including betatron radiation and Compton scattering [13-23].

Extending the distance over which the laser pulse, background plasma, and electron beam interact is critical to increasing either the electron energy gain or X-ray brightness. Three phenomena typically limit the interaction distance: laser pulse diffraction, laser pulse energy depletion, or electron beam slippage from the accelerating phase of the wakefield—dephasing [3,8]. For diffraction in particular, self-guiding, where the transverse ponderomotive force of the laser pulse bores a guiding structure in the plasma, provides one solution. This, however, requires high pulse powers and is often accompanied by rapid depletion of the pulse energy. Preformed plasma waveguides, on the other hand, can guide the pulse over distances unrestrained by vacuum diffraction or depletion [24]. The waveguides (or channels) rely on the refractive index gradient associated with the linear plasma current. This prevents premature depletion of pulse energy, but at the same time, places a limit on the peak pulse amplitudes that can be guided.

While there are several variants of plasma channel structures and formation techniques [24-33], of primary interest here are curved plasma channels [25]. Since the first demonstration by Ehrlich *et al* [25], curved plasma channels have attracted little attention. There are, however, several applications for curved plasma channels. As discussed above, channel guiding for LWFA is only possible for low amplitude pulses. At these amplitudes, the accelerator often requires external injection of electrons. The curved plasma could allow for electron injection and laser pulse coupling into the channel along two different axes: for instance, the electron bunch could be directed along a straight line through the channel walls, while the laser beam is coupled from the side. Furthermore, as we demonstrate below, the wakefields in the curved plasma channel can steer electrons along a curved path. This allows for alternative geometries to the standard linear collider configuration proposed for future laser wakefield accelerators [34]. Perhaps the most obvious application, however, is that electrons accelerated along a curved plasma channel will emit synchrotron radiation.

Here we explore laser pulse guiding, wakefield generation, electron steering, and synchrotron emission in curved plasma channels. We show that a laser pulse guided in a curved plasma channel can excite wakefields that steer electrons along an arched trajectory. The "matched" guided mode is centrifugally displaced from the channel axis, and, as a result, drives an asymmetric wakefield. Distinct from wakefields in a straight plasma channel, the defocusing phase of the wakefield supports stable electron trajectories. These trajectories follow the curve of the channel, while undergoing betatron oscillations centered thereof. Along the trajectory, the electrons emit synchrotron radiation. For experimentally realizable parameters, the radius of curvature corresponds to a magnetic field of ~7 T, such that a ~2 GeV electron emits 0.1 photons per cm with an average photon energy of 25 keV.

**II. Curved plasma channel guiding**

Preformed plasma channels allow for the collimated propagation of high intensity, short laser pulses over distances unrestrained by vacuum diffraction. The guiding persists when the channel is curved, albeit slightly modified. To illustrate the guiding properties of curved plasma channels, we consider a simple paraxial treatment. We express the vector potential of the laser pulse as a plane wave modulating a slowly varying envelope: $\mathbf{A}(\mathbf{x},t) = \tfrac{1}{2}\mathbf{A}_0(\mathbf{x},t)e^{i(kz-\omega t)} + c.c.$, where $\omega$ is the central laser frequency, $k = \omega/c$, and $c$ is the speed of light in vacuum. The low amplitude pulse, $|\mathbf{a}_0| = e|\mathbf{A}_0|/mc < 1$ where $m$ is the electron mass and $e$ the elementary charge, propagates through a tenuous plasma, $k_p^2/k^2 \ll 1$ where $k_p^2 = e^2 n/\varepsilon_0 mc^2$ and $n$ is the electron density. The envelope of the transverse vector potential evolves according to the paraxial wave equation:

$$\left[2ik\frac{\partial}{\partial z}+\nabla_\perp^2\right]\mathbf{a}_\perp(\mathbf{x},\xi) = k_p^2(\mathbf{x})\mathbf{a}_\perp(\mathbf{x},\xi), \quad (1)$$

where $\xi = v_g t - z$ is the coordinate in the axial group velocity, $v_g$, frame of the pulse. The use of Eq. (1) requires that the plasma dispersion and nonlinear response weakly affect the temporal profile of the pulse. As demonstrated elsewhere [3], these requirements can be summarized by the respective inequalities $k_p L \ll (k_p c\tau)^2 (k/k_p)^3$ and $k_p L \ll (k/a_0 k_p)^2$, where $L$ is the propagation distance and $\tau$ the pulse duration. With a solution of Eq. (1), the approximate axial vector potential can be found from $\nabla \cdot \mathbf{A}(\mathbf{x},t) \approx 0$. For the parameters considered here, however, $\mathbf{a}_0 \approx \mathbf{a}_\perp$.

The electron density profile for the curved plasma channel can be modeled as

$$n(\mathbf{x}) = n_0 + \frac{1}{2}n_0''[x - x_A(z)]^2 + \frac{1}{2}n_0'' y^2, \quad (2)$$

where the density has a minimum of $n_0$ along the parabolic, parametric curve $x = x_A \equiv (z - z_n)^2 / 2R$ and $y = 0$. The curve, which we also refer to as the channel axis (subscript A), represents a local expansion about a circular path in the $x-z$ plane centered at $x = R$ and $z = z_n$ with a radius $R$. The coefficient $n_0''$ determines the width of the channel. Figure (1) displays an example of the density profile when $z_n = 0$. In order for Eqs. (1) and (2) to be consistent within the paraxial approximation $(L/R)^2 \ll 1$. This is not a physical limitation on $L$. Over longer distances, including full circular paths, one could patch together Eqs. (1) and (2) over multiple arcs or switch to toroidal coordinates with z replaced by $R\phi$ where $\phi$ is the toroidal angle. Such a treatment is outside the scope of the current manuscript.

Equations (1) and (2) permit analytical solutions for the evolution of transverse Hermite-Gaussian pulses in a curved plasma channel. For the lowest order mode, we write

$$\mathbf{a}_0(\mathbf{x},\xi) = \hat{\mathbf{a}}_0(\xi)\exp\left[i\mathbf{k}_c \cdot (\mathbf{r}-\mathbf{r}_c) - \sum_{q=x,y}(1+i\alpha_q)\frac{(q-q_c)^2}{w_q^2} + i\theta\right], \quad (3)$$

where $\mathbf{k}_c$ is the transverse wavenumber, $\mathbf{r}$ the transverse coordinate, $\mathbf{r}_c$ the centroid displacement, $\alpha_q$ the phase curvature coefficient, $w_q$ the spot size, and $\theta$ the phase, each of which are real functions of $z$. One can obtain a set of linear differential equations for these quantities by substituting Eqs. (2) and (3) into Eq. (1), matching like powers of transverse coordinates, and simplifying. In general, the solution for $\mathbf{r}_c$ oscillates about the density minimum of the channel with a period $\lambda_r = \pi k w_m^2$, where $w_m = (e^2 n_0'' / 8\varepsilon_0 mc^2)^{-1/4}$ is the matched spot size. The oscillation amplitude is determined by the initial centroid offset and transverse wavenumber, $\mathbf{k}_c = k \partial_z \mathbf{r}_c$. Similarly, $w_q$ undergoes oscillations about its initial value with a period $\lambda_w = \frac{1}{2} \pi k w_m^2$ and an amplitude determined by the initial conditions of the phase curvature coefficient, $\alpha_q = -\frac{1}{4} k \partial_z w_q^2$.

Here we limit the investigation to "matched" solutions for which neither the centroid displacement nor spot size undergo oscillations. The solutions can be expressed as

$$x_c = x_A - \frac{Z_r^2}{R} \quad (4)$$

$k_x = k(z - z_n)/R$, $y_c = k_y = 0$, $\alpha_q = 0$, $w_q = w_m$, and

$$\theta = -\left( \frac{1}{Z_R} + \frac{k_{p0}^2}{2k} + \frac{kZ_r^2}{2R^2} \right) z + \frac{k}{6R^2}(z - z_n)^3,$$

where $Z_r = \frac{1}{2} k w_m^2$ and $k_{p0} = (e^2 n_0 / \varepsilon_0 mc^2)^{1/2}$. Equation (4) predicts that, for matched channel propagation, the laser pulse must be coupled into the channel with an offset from the channel axis, $x_c = (z_n^2 / 2R) - (Z_r^2 / R)$ at $z = 0$. Once coupled into the channel, the centroid follows the parabolic trajectory of the electron density minimum with a small centrifugal displacement to the outside edge of the channel, the first and second terms in Eq. (4) respectively. The curvature of the channel also modifies the local phase velocity. Specifically, $v_{pz} \simeq v_{p0} + c[\frac{1}{2}(Z_r / R)^2 - (x_A / R)]$ and $v_{px} \simeq -c(z - z_n)/R$, where $v_{p0} = c[1+(kZ_r)^{-1}+\frac{1}{2}(k_{p0}/k)^2]$ is the uniform channel phase velocity [30]. The phase fronts of the guided mode remain orthogonal to the curve of the channel.

## III. Wakefields in a curved channel

The laser pulse ponderomotively excites wakefields as it propagates through the curved plasma channel. Using a separation of time scales based on the disparity between the laser pulse and plasma frequencies, the equation for the wakefields in a non-uniform plasma can be found from the fluid and Maxwell's equations [35]:

$$\left[\frac{\partial^2}{\partial \xi^2}+k_p^2(\mathbf{x})\right]\mathbf{E} = -\frac{mc^2}{4e}k_p^2(\mathbf{x})(\nabla_\perp - \hat{\mathbf{z}}\partial_\xi)|\mathbf{a}_0|^2. \quad (5)$$

For a pulse with the temporal profile $|\hat{\mathbf{a}}_0(\xi)|^2 = a_0^2 \sin^2(k_\tau \xi/2)$ on the domain $0 < \xi < 2\pi/k_\tau$, the wakefields behind the pulse, $\xi \geq 2\pi/k_\tau$, are given by

$$E_x = -\frac{mc^2 a_0^2}{2ew_m^2}\left[\frac{(x-x_c)f(\mathbf{x})}{\eta^2-1}\right]\left[\sin(2\pi\eta)\sin(k_p\xi) - 2\sin^2(\pi\eta)\cos(k_p\xi)\right] \quad (6a)$$

$$E_y = -\frac{mc^2 a_0^2}{2ew_m^2}\left[\frac{yf(\mathbf{x})}{\eta^2-1}\right]\left[\sin(2\pi\eta)\sin(k_p\xi) - 2\sin^2(\pi\eta)\cos(k_p\xi)\right] \quad (6b)$$

$$E_z = -\frac{mc^2 k_\tau a_0^2}{8e}\left[\frac{\eta f(\mathbf{x})}{\eta^2-1}\right]\left[\sin(2\pi\eta)\cos(k_p\xi) + 2\sin^2(\pi\eta)\sin(k_p\xi)\right] \quad (6c)$$

where we have defined $\eta = k_p(\mathbf{x})/k_\tau$ and $f(\mathbf{x}) = \exp[-2w_m^{-2}(x-x_c)^2 - 2w_m^{-2}y^2]$.

The wakefields in a curved plasma channel have several distinct features. These can be observed in Fig. (2), which displays $eE_x/k_{p0}mc^2$, top, and $eE_z/k_{p0}mc^2$, bottom, behind the pulse at $y = 0$ and $k_{p0}z = 19$ as a function of $x$ and $\xi$ for the parameters found in Table I. The on-axis peak wakefield amplitude was maximized by setting the pulse full width at half maximum (FWHM) duration $\tau = \pi/ck_\tau$ equal to $\pi/ck_{p0}$. The peak intensity of the laser pulse occurs slightly off axis, however, causing an obvious asymmetry. Specifically, the horizontal null in $eE_x/k_{p0}mc^2$ and extrema of $eE_z/k_{p0}mc^2$ are colocated with the transverse position of the laser pulse centroid at $k_{p0}x = -0.15$, while the channel axis is at $x = x_A \approx 0$. From the channel axis outward, the plasma period decreases, bowing the plasma wave phase fronts and rapidly phase-mixing the electrostatic fields.

The centrifugal offset of the channel axis and centroid, $x_A - x_c = Z_r^2 / R$, results in $E_x$ having a larger peak amplitude above the centroid than below. Said differently, $|E_x|$ has a single maximum at a focusing (or defocusing) phase. This is in contrast to an axially uniform plasma in which $|E_x|$ has two maxima at a focusing (or defocusing) phase, located symmetrically about the centroid at $x = \pm \frac{1}{2} w_m$. To show this explicitly, we consider Eq. (6a) near a focusing phase to lowest order in the plasma wavenumber spatial variation:

$$E_x = -\frac{\pi m c^2 a_0^2}{2 e w_m^2}(x - x_c) f(\mathbf{x}) \left[ 1 - \frac{(x - x_A)^2}{k_{p0}^2 w_m^4} \right]. \quad (7)$$

In a uniform plasma, $x_A = x_c = 0$ and the second term in square brackets is equal for $x = \pm \frac{1}{2} w_m$. In the curved channel, on the other hand, we set $x = x_c \pm \frac{1}{2} w_m$ and find $E_x \propto 1 - [(w_m \Box 2R^{-1} Z_r^2)/2 k_{p0} w_m^2]^2$. As expected, the field is larger above the centroid axis (-) than below (+). The axial field, while asymmetric, peaks near the centroid axis, $x \approx x_c$, similar to a uniform plasma.

**IV. Synchrotron motion**

Relativistic electrons trailing the laser pulse will evolve in response to the wakefields. The transverse wakefields, in particular, can steer electrons along the curved path of the channel. To demonstrate this, we perform test particle simulations. The simulation evolves the electron equations of motion, $\dot{\mathbf{x}} = \mathbf{p}/m\gamma$ and $\dot{\mathbf{p}} = -e\mathbf{E}$ where $\gamma = [1 + (p/mc)^2]^{1/2}$ is the Lorentz factor, using the fields in Eq. (6). Radiation losses, which we consider later, are small and thus neglected. The electrons are initialized with $y = p_x = p_y = 0$ and an axial velocity $v_z = c(1 - \gamma_0^{-2})^{1/2}$. The remaining laser pulse, plasma, and electron bunch parameters can be found in Table I unless otherwise stated.

Figure (3a) shows the initial conditions that end within $\pm w_m$ of the channel axis overlayed on the initial transverse wakefield, while (3b) shows the corresponding trajectories overlayed on the background electron density. Only those electrons starting in defocusing phases above the centroid (yellow) or focusing phases below the centroid (blue) remain in the channel, confirming that the transverse fields steer the electrons along the curved path. A larger spread of positions starting above the centroid axis than below remain confined to the channel, consistent with the asymmetry in the transverse field amplitudes discussed above.

For a more general description of the electron dynamics, we obtain approximate solutions to the equations of motion. Assuming weak spatial variation of the plasma density, the equations of motion reduce to

$$\frac{dp_z}{dt} = \frac{\pi m c^2 k_{p0} a_0^2}{8} f(x) \cos(k_{p0}\xi) \quad (8a)$$

$$\frac{dp_x}{dt} = \frac{\pi m c^2 a_0^2}{2 w_m^2}[x - x_c(z)] f(x) \sin(k_{p0}\xi) \, , \quad (8b)$$

and $y = p_y = 0$. Equation (8) can be further simplified by performing an expansion in inverse powers of the initial Lorentz factor, $\gamma_0 = (1 - v_0^2/c^2)^{-1/2}$. To second order, we have $x(t) = x_1(t)$ and $z(t) = z_i + v_0 t + z_2(t)$, which evolve according to

$$\frac{d^2 x_1}{dt^2} = -\Omega^2 [x_1 - x_c(z_i + v_0 t)] f(x_1) \sin(k_{p0} z_i) \quad (9a)$$

$$\frac{d^2 z_2}{dt^2} \simeq -\frac{1}{2c}\frac{d}{dt}\left(\frac{dx_1}{dt}\right)^2 , \quad (9b)$$

where the numbered subscripts denote the perturbation order, $z_i$ is the initial axial position, $\Omega = (c^2 \pi a_0^2 / 2\gamma_0 w_m^2)^{1/2}$ is the betatron frequency for a weakly nonlinear wakefield, and we have assumed distances much shorter than a dephasing length, $k_{p0} L (v_0 - v_g)/c \ll 1$, consistent with Table I.

As observed in Fig. (3), the electrons that remain confined to channel start in either a defocusing or focusing phase. Taylor expanding Eq. (9a) about $k_{p0}z_i = -5\pi/2$ and $k_{p0}z_i = -7\pi/2$ for the defocusing (+) and focusing phases (-) respectively, we have

$$x_1 = x_i + \frac{c^2 t^2}{2R} + \chi \quad (10)$$

$$\frac{d^2\chi}{dt^2} = -\frac{c^2}{R} \pm \Omega^2 \left(x_i + \frac{Z_r^2}{R} + \chi\right) \exp\left[-\frac{2}{w_m^2}\left(x_i + \frac{Z_r^2}{R} + \chi\right)^2\right], \quad (11)$$

where $\chi$ represents the deviation of the electron trajectory from the curve of the channel. Equations (10) and (11) demonstrate that in the absence of the transverse wakefield forces, $\Omega \to 0$, the electrons would follow a straight path, $x_1 = x_i$, and quickly leave the channel. With the transverse forces, however, the electrons can remain in the channel over an extended distance.

"Ideal" electron trajectories, in particular, have $\chi = 0$ and exactly track the curve of the channel, $x_1 = x_i + c^2 t^2 / 2R$. The initial conditions for the ideal trajectories can be found be setting $\chi = 0$ in Eq. (11): $x_i^\circ = -(Z_r^2/R) \pm \frac{1}{2} w_m [-W_0(\alpha)]^{1/2}$ and $x_i^\square = -(Z_r^2/R) \pm \frac{1}{2} w_m [-W_{-1}(\alpha)]^{1/2}$ where $W_0$ and $W_{-1}$ are the upper and lower branches of the Lambert W function for real argument and

$$\alpha = -\left(\frac{4\gamma_0 w_m}{\pi R a_0^2}\right)^2. \quad (12)$$

The Lambert W function is real, negative for $-\exp(-1) < \alpha < 0$. The value of $\alpha$ therefore determines whether the laser pulse, channel, and relativistic electron interaction will support ideal trajectories. The $\alpha \approx 0$ limit corresponds to strong transverse wakefields or weak channel curvature, while the $\alpha \approx -\exp(-1)$ limit has the opposite correspondence. For even smaller values of $\alpha$, the transverse wakefields are not strong enough to confine relativistic electrons to the channel indefinitely. Ignoring the centrifugal term $Z_r^2/R$ for the moment, the $\pm$ signs in the expression for $x_i^\circ$ and $x_i^\square$ predict the behavior observed in the simulation: the transverse wakefields need to push the electron in the direction that the channel curves. The centrifugal term slightly breaks the symmetry. When $\alpha = -\exp(-1)$, $x_i^\circ = x_i^\bullet \approx x_c \pm \frac{1}{2}w_m$, which are the approximate locations of the strongest defocusing and focusing fields.

In Fig. (3), $\alpha$ was equal to $-0.17$ and large swaths of initial conditions were confined to the channel. For comparison, Figs. (4a) and (4b) display the confined initial conditions for $\alpha = -0.37 \approx -\exp(-1)$ and $\alpha = -0.44$ respectively. The value of $\alpha$ was modified by changing $w_m$, keeping all other parameters identical to Table I; the values $\alpha = -0.37$ and $\alpha = -0.44$ differ by only 10% in the spot size. The reduction in the region of confined initial conditions is apparent in both Figs. (4a) and (4b). When $\alpha = -0.37$ the focusing phase is unable to confine electrons to channel, while when $\alpha = -0.44$ neither the focusing nor the defocusing phase confines the electrons. The larger spot sizes weaken the transverse ponderomotive force and resulting wakefield, diminishing the fields ability to steer electrons along the channel. This qualitative difference between the two channels is surprising considering the wakefields appear almost identical.

While the existence of the ideal trajectory provides a useful condition on the laser pulse and channel parameters, matching the electrons' initial condition to $x_i = x_i^\circ$ and $x_i = x_i^\square$ is unrealistically restrictive. We would like to know, therefore, how sensitive the electron trajectories are to slight deviations, $\delta x_i^{\circ,\bullet} = x_i - x_i^{\circ,\bullet}$, from the ideal initial conditions. By expanding Eq. (11) for small $\delta x_i^{\circ,\bullet}$, we can determine how quickly non-ideal electrons deviate from the curve of the channel. Upon going through the algebra, we have

$$\chi^\circ \simeq -2\delta x_i^\circ \sinh^2(\sqrt{\pm \tfrac{1}{2}}\hat{\Omega}_0 t) \quad (13a)$$

$$\chi^\bullet \simeq 2\delta x_i^\bullet \sin^2(\sqrt{\pm \tfrac{1}{2}}\hat{\Omega}_{-1} t) \quad (13b)$$

where $\hat{\Omega}_0 = [1 + W_0(\alpha)]^{1/2} \exp[\tfrac{1}{4}W_0(\alpha)]\Omega$ and $\hat{\Omega}_{-1} = [|W_{-1}(\alpha)| - 1]^{1/2} \exp[\tfrac{1}{4}W_{-1}(\alpha)]\Omega$. Equations (13) demonstrate that stable solutions for the electron motion exist in both the focusing and defocusing phases, $\chi^\circ$ and $\chi^\square$ respectively. Specifically, the electrons undergo small amplitude betatron oscillations about the curve of the channel. To within the approximations made here, a stable trajectory does *not* exist in the defocusing phase

of a straight channel wakefield. In the limit of a straight channel, $R \to \infty$, $\alpha \to 0$, and $x^\square \to \infty$: the stable point in the defocusing phase is at infinity. Stable trajectories in the defocusing phase are a distinct feature of curved channels that allow for electron steering.

The analysis above neglects the spatial variation in the plasma wavenumber. As discussed in connection with Fig. (2), the variation produces an asymmetry in the wakefields. In particular, the amplitude of the wakefield in the defocusing phase for $x > x_c$ is larger than in the focusing phase for $x < x_c$. The result is that the defocusing phase steers a larger spread of initial conditions along the curve of the channel. In addition, the exact values of $x_i^\circ$ and $x_i^\square$ will be slightly modified by the spatial variation in the plasma wavenumber. The analysis, nonetheless, illustrates the underlying process that confines electrons to the channel.

With the above considerations, we can approximate the trajectories of electrons that remain in the channel as

$$x_1 \simeq x_i + \frac{c^2 t^2}{2R} \quad (14a)$$

$$z_2 \simeq -\frac{c^3 t^3}{6R^2}, \quad (14b)$$

which represent synchrotron trajectories for an arc with radius of curvature equal to that of the channel, $R$. Figures (5a) and (5b) show a comparison of Eqs. (14a) and (15b) with a randomly selected confined electron trajectory from the test particle simulation. The agreement is apparent.

**V. Synchrotron radiation**

As the electrons are accelerated along the curve of the channel, they emit synchrotron radiation. The power radiated by a single electron can be found from the Larmor formula [36]

$$P_s(t) = \frac{e^2}{6\pi\varepsilon_0 c} \gamma^6 \left[ (\dot{\boldsymbol{\beta}})^2 - (\boldsymbol{\beta} \times \dot{\boldsymbol{\beta}})^2 \right], \quad (15)$$

where $\boldsymbol{\beta}$ is the velocity normalized to $c$. With use of Eqs. (14), this simplifies to the usual expression for synchrotron radiation, $P_s(t) = ce^2 \gamma_0^4 / 6\pi\varepsilon_0 R^2$, examples of which can be found in Table I. Noting that the axial momentum contributes most of the kinetic energy, the radiation reaction (damping) can be approximated as $\dot{p}_z |_{rad} = -p_z P_s / mc^2$ with the associated damping length $L_{rad} = (3/2)\gamma_0^{-4} r_e^{-1} R^2$, where $r_e$ is the classical electron radius. In both examples presented in Table I, the damping length far exceeds the distances considered, justifying the absence of radiation damping in the analysis and simulations.

Of more interest for applications is the forward radiated spectrum. The radiated energy per unit frequency per unit solid angle, $\Omega$, can be calculated from

$$\frac{d^2 U}{d\omega d\Omega} = \frac{e^2 \omega^2}{16\pi^3 \varepsilon_0 c} \left| \int dt \left[ \mathbf{n} \times (\mathbf{n} \times \boldsymbol{\beta}) \right] \exp\left[ i\omega(t - \mathbf{n} \cdot \mathbf{x}/c) \right] \right|^2, \quad (16)$$

where $\mathbf{n}$ is the unit vector pointing to the location of observation [36]. As demonstrated elsewhere [36], the radiation is emitted predominately in the forward direction and in the plane of motion. When integrated over all time, the resulting on-axis spectrum is

$$\left. \frac{d^2 U}{d\omega d\Omega} \right|_{\theta=0} = \frac{e^2 \gamma_0^2}{4\pi^3 \varepsilon_0 c} \psi^2 K_{2/3}^2(\psi), \quad (17)$$

where $\psi = \omega / 2\omega_s$, $\omega_s = 3c\gamma_0^3 / 2R$ is the characteristic synchrotron radiation frequency, and $\theta$ defines the angle between the emission and momentum directions. Figure (6) displays a comparison of Eq. (17) with numerical calculations of Eq. (16) using Eq. (14) and a confined electron trajectory from the test particle simulation. For the numerical calculations, the integrand of Eq. (16) was approximated as $\beta_x \exp[i\omega(t - z/c)]$, consistent with the derivation of Eq. (17) and small angle emission. The resulting spectrum was Gaussian filtered for clarity in presentation. The figure shows reasonable agreement between the three curves. The simulated spectra were integrated over the finite duration of the simulation, explaining the discrepancy with Eq. (17). The betatron motion, captured only in the test particle simulation, additionally modifies the spectra.

The characteristic frequency depends strongly on the electron energy, $\omega_s \propto \gamma_0^3$. This dependence and recently achieved multi-GeV LWFA electrons [4-6] motivated our use $\gamma_0 = 4000$ in the analysis above. Given a high-energy electron source, the utility of a curved plasma channel synchrotron ultimately relies on the number of radiated photons.

We can estimate the number of photons as $N_p \approx N_e P_s L / c \langle \hbar \omega \rangle$ where $N_e$ is the number of electrons undergoing synchrotron motion and $\langle \hbar \omega \rangle = U^{-1} \int \hbar \omega (dU/d\omega) d\omega$ is the average photon energy. Approximating $\frac{d^2 U}{d\omega d\Omega} \simeq \frac{d^2 U}{d\omega d\Omega}|_{\theta=0} \exp(-3\omega \gamma_0^2 \theta^2 / 2\omega_s)$ as in Ref [36] and integrating over angle and frequency, we find $\langle \hbar \omega \rangle \approx 1.3 \hbar \omega_s$ and

$$N_p \approx 0.8 \left( \frac{e^2}{9\pi \hbar \varepsilon_0 c} \right) \left( \frac{N_e \gamma_0 L}{R} \right). \quad (18)$$

The total number of photons increases with the channel length and electron energy, and decreases with the channel curvature. For both cases in Table I, Eq. (18) evaluates to $N_p = 0.3 N_e$.

### VI. Practical considerations

For an experimental realization, one would need to create a pre-formed curved plasma channel. One possibility is to use a heater pulse, for instance a ~100ps Nd:YAG pulse, with an Airy profile. The peak intensity of an Airy beam follows a parabolic trajectory [37,38]. When incident on a gas jet, the beam would ionize the gas and heat the plasma along that trajectory. The resulting hydrodynamic expansion would provide the curved plasma channel. A second option would be to manufacture a curved capillary as in Ref [25]. The profile of the discharge plasma would reflect the capillary geometry. An alternative to both of these is the recently proposed long plasma channel concept based on colliding gas flows [31]. The collision of the flows sustains an on-axis neutral density minimum, which can be subsequently ionized by a discharge or laser pulse. The scheme can be implemented in a curved geometry by guiding the flow along a curved transparent tube.

In addition to compactness, the small radius of curvature of the curved plasma channel provides an increase in the synchrotron radiated power and favors higher emission frequencies when compared to full-size synchrotrons [39]. Furthermore, with the recently achieved multi-GeV electrons from LWFA, the electron energies are comparable to those used in the National Synchrotron Light Source II (NSLS-II), ~3 GeV [39]. The total radiated energy, however, is far smaller on account of the lower charge,

~10 pC compared to ~1 nC for NSLS-II [39]. As a rough guideline, the peak ratio of $L/R$ would be comparable in both cases to ensure forward emission.

In principle, the maximum steerable charge could be limited by beam loading. For a relativistic electron bunch, however, the transverse space charge force is typically quite small: the Lorentz force of the azimuthal magnetic field generated by the current nearly cancels the electrostatic repulsion. To estimate the bunch density at which beam loading affects the steering, $n_b$, we can compare the transverse wakefield and space charge forces. Approximating the beam as a uniform density cylinder, and forming the ratio of the resulting force with that of the wakefield, we find $n_b = \gamma_0^2 a_0^2 / 4\pi r_e w_m R_b$, where $R_b$ is the bunch radius. For the parameters in Table I and any reasonable choice of $R_b$, this density exceeds, by several orders of magnitude, those found in typical electron bunches.

As an alternative method for achieving a small radius of curvature, one may suggest a dipole magnet. For the purpose of comparison, we can define the effective magnetic field of the curved plasma channel by equating the radius of curvature to the Larmor radius: $B_{eff} = mc\gamma_0 / eR$. In the second and third columns of Table I, $B_{eff} = 6.8$ T and $B_{eff} = 11.8$ T respectively, while the superconducting magnets of the Large Hadron Collider (LHC) operate at 8 T [40]. The corresponding magnetic fields are not readily available in typical laser-plasma interaction laboratories.

**VII. Summary and conclusions**

We have examined laser pulse guiding, wakefield generation, electron steering, and synchrotron emission in curved plasma channels. The matching conditions for the guided mode were presented and revealed that the centroid is centrifugally displaced to the outer edge of the channel. The resulting wakefields exhibit several features distinct from the case of matched propagation in a straight plasma channel. Foremost, the transverse wakefield has only a single global extremum in a focusing or defocusing phase with a larger amplitude along the inner edge of the channel. Second, the peak axial wakefield is displaced transversely from the channel axis. Finally, and most importantly, both the focusing and defocusing phases of the wakefield can support stable, relativistic electron trajectories.

The stability of the focusing and defocusing phases relies on the existence of "ideal" initial conditions. Electrons starting at an ideal initial condition remain confined to the channel indefinitely, exactly following the channel curve. Electrons starting near a stable ideal initial condition also follow the channel curve, but undergo betatron oscillations centered thereof. The dimensionless parameter $\alpha$, which combines laser pulse, plasma channel, and electron beam parameters, was introduced and indicates whether or not the transverse wakefields can steer the electrons along the channel.

As the electrons are steered along the channel, they emit synchrotron radiation. A synchrotron source based on a curved plasma channel LWFA may have an advantage in terms of compactness and its non-reliance on high magnetic field technology. In addition, the small radius of curvature of the channel favors higher powers and emission frequencies when compared to large-scale synchrotron devices. The total radiated energy, however, is far smaller on account of the lower charge [39]. For experimentally realizable laser and plasma parameters, a ~2 GeV electron emits 0.1 photons per cm with an average photon energy of multiple keV.

**Acknowledgements**
This work was supported by the Naval Research Laboratory 6.1 Base Program.

| Parameter | Normalized value | Value when $n_0 = 1 \times 10^{17}$ cm$^{-3}$ | Value when $n_0 = 3 \times 10^{17}$ cm$^{-3}$ |
|---|---|---|---|
| $a_0$ | 0.5 | 0.5 | 0.5 |
| $\tau$ | $\pi$ | 180 fs | 100 fs |
| $w_m$ | 1.2 | 20 μm | 9.6 μm |
| $z_n$ | 0 | 0 | 0 |
| $R$ | $5.95 \times 10^4$ | 1.0 m | 58 cm |
| $L$ | 1340 | 2.25 cm | 1.3 cm |
| $\gamma_0$ | 4000 | 2 GeV | 2 GeV |
| $\alpha$ | -.17 | -.17 | -.17 |
| $P_s$ | - | 74 eV/ps | 220 eV/ps |
| $L_{rad}$ | $1.2 \times 10^5$ | 2.0 m | 0.7 m |
| $\hbar\omega_s$ | - | 19 keV | 33 keV |
| $B_{eff}$ | - | 6.8 T | 11.8 T |

Table 1. Parameters for calculations and simulations unless otherwise stated. Normalized distances and times are multiplied by $k_{p0}$ and $ck_{p0}$ respectively.

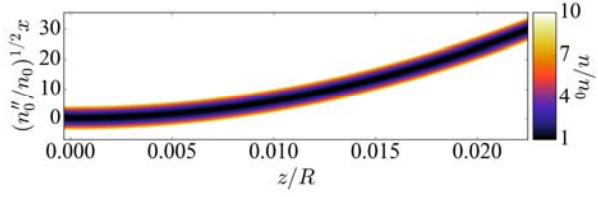

Figure 1. Plasma density of a curved plasma channel as a function of the transverse and axial coordinates. The color scale is saturated at 10 times the on-axis density. The apparent narrowing of the channel for large z is an artifact of the aspect ratio.

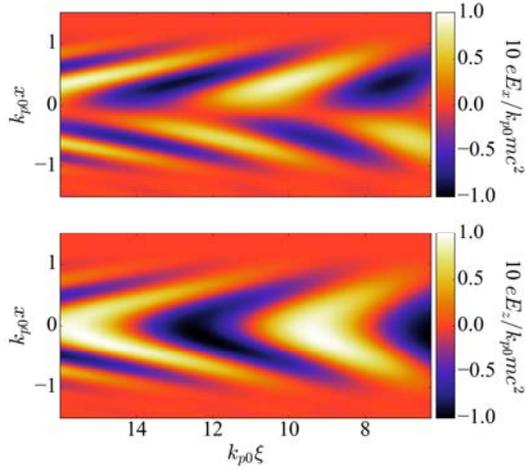

Figure 2. Transverse and longitudinal wakefields behind the laser pulse as a function of the transverse and pulse frame coordinates.

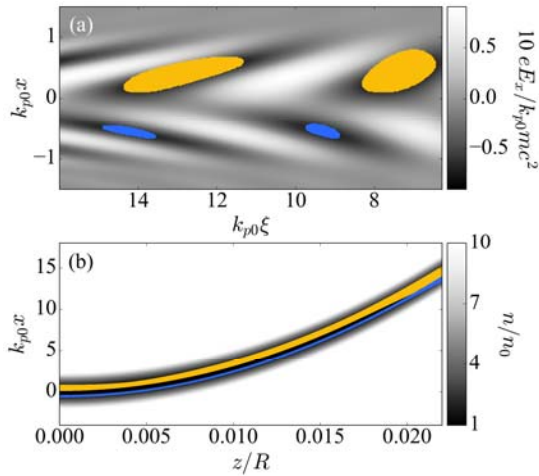

Figure 3. (a) Electron initial conditions that stay within $\pm w_m$ of the channel overlayed on the initial transverse wakefield. Initial conditions in the top (yellow) swaths start in defocusing phases while the bottom (blue) start in focusing phases. (b) The trajectories

corresponding to the initial conditions in (a) overlayed on the background electron density.

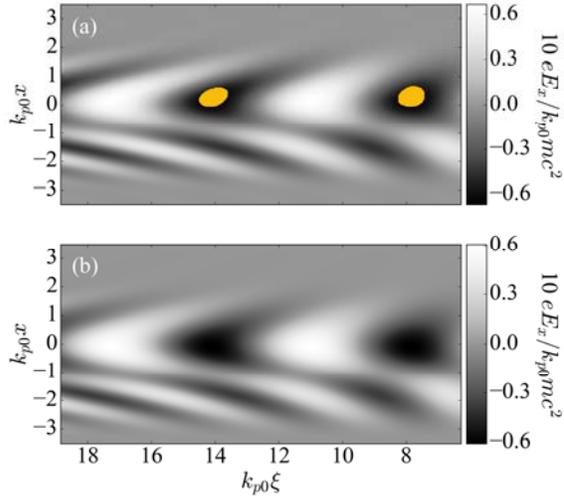

Figure 4. Electron initial conditions that stay within $\pm w_m$ of the channel overlayed on the initial transverse wakefield: (a) $\alpha = -.37$ and (b) $\alpha = -.44$.

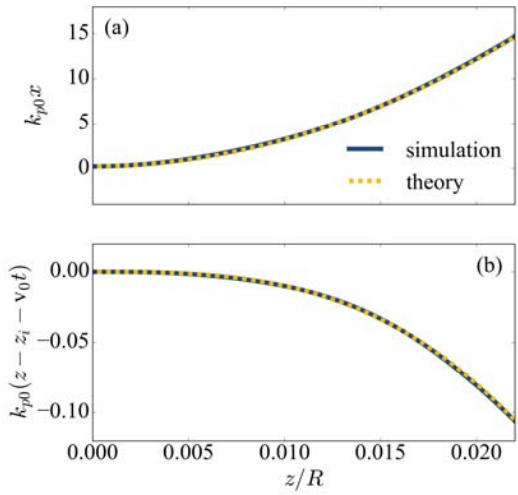

Figure 5. Comparison of Eqs. (14a) and (15b) with a randomly selected confined electron trajectory from the test particle simulation.

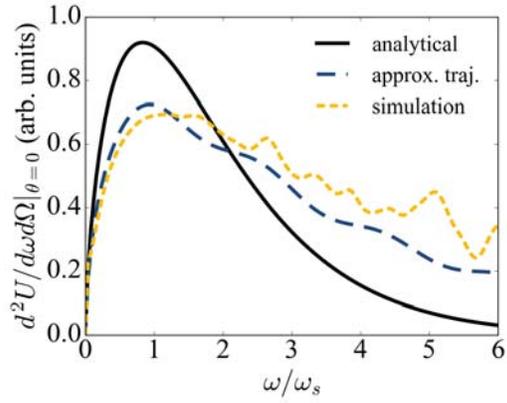

Figure 6. Comparison of the small angle synchrotron radiation spectra from the analytical theory, calculated using the approximate trajectories, and calculated using a trajectory from the test particle simulation.